\documentclass[reprint, aps, amsmath, amssymb]{revtex4-2}

\usepackage{graphicx}
\usepackage{dcolumn}
\usepackage{bm}
\usepackage{siunitx}
\usepackage{multirow}
\usepackage[mathlines]{lineno}
\usepackage{orcidlink}
\DeclareUnicodeCharacter{2212}{-}
\begin{document}
\preprint{APS/123-QED}
\title{Search for $h_b(2P)\to\gamma\chi_{bJ}(1P)$ at $\sqrt{s} = 10.860$ GeV}
\noaffiliation
  \author{A.~Boschetti\,\orcidlink{0000-0001-6030-3087}} 
  \author{R.~Mussa\,\orcidlink{0000-0002-0294-9071}} 
  \author{U.~Tamponi\,\orcidlink{0000-0001-6651-0706}} 
  \author{I.~Adachi\,\orcidlink{0000-0003-2287-0173}} 
  \author{H.~Aihara\,\orcidlink{0000-0002-1907-5964}} 
  \author{D.~M.~Asner\,\orcidlink{0000-0002-1586-5790}} 
  \author{T.~Aushev\,\orcidlink{0000-0002-6347-7055}} 
  \author{R.~Ayad\,\orcidlink{0000-0003-3466-9290}} 
  \author{Sw.~Banerjee\,\orcidlink{0000-0001-8852-2409}} 
  \author{K.~Belous\,\orcidlink{0000-0003-0014-2589}} 
% ...
\collaboration{The Belle Collaboration}

\begin{abstract}
In the bottomonium sector, the hindered magnetic dipole (M1) transitions between P-wave states $h_b(2P) \rightarrow  \chi_{bJ}(1P) \gamma$, $J=0, \, 1, \, 2$, are expected to be severely suppressed according to the Relativized Quark Model, due to the spin flip of the $b$ quark. Nevertheless, a recent model following the coupled-channel approach predicts the corresponding branching fractions to be enhanced by orders of magnitude. In this Letter, we report the first search for such transitions. 
We find no significant signals and set upper limits at 90\% CL on the corresponding branching fractions: $\mathcal{B}[h_b(2P)\to\gamma\chi_{b0}(1P)] < 2.7 \times 10^{-1}$, $\mathcal{B}[h_b(2P)\to\gamma\chi_{b1}(1P)] < 5.4 \times 10^{-3}$ and $\mathcal{B}[h_b(2P)\to\gamma\chi_{b2}(1P)] < 1.3 \times 10^{-2}$. These values help to constrain the parameters of the coupled-channel models. 
The results are obtained using a \qty{121.4}{fb^{-1}} data sample taken around $\sqrt{s}=$ \qty{10.860}{GeV} with the Belle detector at the KEKB asymmetric-energy $e^+e^-$ collider.

\end{abstract}
\maketitle
    In recent years, bottomonium spectroscopy has shown a number of unexpected results. A key one among them is the observation of the spin-singlet P-wave state $h_b(2P)$ by the Belle collaboration \cite{Belle:2011wqq}. The production of spin-singlet bottomonia is generally rare in $e^+e^-$ collisions because it requires the spin flip of a heavy quark in a hadronic or radiative transition. Nevertheless, $h_b(2P)$ is produced via the $\Upsilon(10860)\rightarrow h_b(2P) \pi^+\pi^-$ transition with a surprisingly large rate. The observation of this unexpected enhancement has led to the discovery of the exotic four-quark states $Z_b(10610)^\pm$ and $Z_b(10650)^\pm$ \cite{Belle:2011aa} that are produced as intermediate states in the dipion transitions.

    In this analysis, we search for the hindered magnetic dipole (M1) transitions between the spin-singlet and spin-triplet states $h_b(2P) \to \gamma \chi_{bJ}(1P)$. This is the first search for such kind of transitions. According to the Relativized Quark Model \cite{Godfrey:1985xj}, these are expected to be severely suppressed because of the heavy quark spin flip, and branching fractions of order $10^{-6} - 10^{-5}$ are predicted. Nevertheless, a recent study \cite{guo2016} considering coupled-channel effects giving rise to $B$ meson loop diagrams predicts the branching fractions for the considered transitions to be in the order of $10^{-2} - 10^{-1}$. The loop diagrams involving the production of virtual $B$ mesons have been used to model other hidden-bottom transitions, such as $\Upsilon(3S)\to\pi^0 h_b(1P)$, whose observation has been recently claimed by BaBar \cite{BaBar:2011ljf}. Therefore, experimental results are needed for a correct understanding of the transitions. 
    Many recent results in the quarkonium sector show that transitions which were once unlikely to observe are now accessible. We are thus further motivated to search for the hindered M1 transitions between P-wave bottomonia for the first time.

    The analysis setup is fully exclusive. We reconstruct the dipion transition from $\Upsilon(10860)$ in order to tag the $h_b(2P)$ production. The $\chi_{bJ}(1P)$ have large radiative decay to the $\Upsilon(1S)$, which can subsequently decay to $\mu^+\mu^-$. The radiative decays $\chi_{bJ}(1P) \to \gamma \Upsilon(1S)$ are thus chosen to tag the final state of the M1 transition, so that there are two photons in the cascade. Therefore, we reconstruct the complete decay chain  $\Upsilon(10860)\rightarrow h_b(2P) \pi^+\pi^- \to \gamma \chi_{bJ}(1P) \pi^+\pi^- \to \gamma \gamma \Upsilon(1S) \pi^+\pi^-  \to \gamma \gamma \mu^+\mu^- \pi^+\pi^-$. 
    We label as $\gamma_1$ the photon radiated in the $h_b(2P)\to\gamma\chi_{bJ}(1P)$ decay, and $\gamma_2$ the photon emitted in $\chi_{bJ}(1P)\to\gamma\Upsilon(1S)$ decays.

    We use a \qty{121.4}{fb^{-1}} data sample collected at the $\Upsilon(10860)$ resonance by the Belle detector \cite{belle2002belle, Belle:2012iwr} at the KEKB asymmetric-energy $e^+e^-$ collider \cite{Kurokawa:2001nw, Abe:2013kxa}. The average center-of-mass (c.m.) energy of this sample is $\sqrt{s}$ = \qty{10.866}{GeV}. The Belle detector was a large-solid-angle magnetic spectrometer that consisted of a silicon vertex detector, a 50-layer central drift chamber (CDC), an array of aerogel threshold Cherenkov counters (ACC), a barrel-like arrangement of time-of-flight scintillation counters, and an electromagnetic calorimeter comprised of CsI(Tl) crystals (ECL) located inside a superconducting solenoid coil that provided a 1.5 T magnetic field. An iron flux-return yoke located outside of the coil (KLM) was instrumented with resistive-plate chambers to detect $K^0_L$ mesons and muons.

    For the Monte Carlo (MC) simulation, we use the EvtGen generator \cite{Lange:2001uf}. Following the results presented in Ref. \cite{Belle:2011wqq}, the $\Upsilon(10860) \to h_b(2P) \pi^+\pi^−$ transition is simulated assuming it proceeds exclusively through the intermediate $Z_b(10610, 10650)^\pm$ states. Each angular distribution in the signal decay chain is generated according to the corresponding spin dynamics. Final-state radiation is included using the PHOTOS package \cite{Davidson:2010ew}. Background MC samples include the production of $B^+$, $B^0$, and $B_s^0$ mesons; the continuum processes $e^+e^- \to q \overline{q}$, $q=u,d,s,c$; and two-photon interactions. The detector response is modeled using GEANT3 \cite{Brun:1119728}. Simulation takes into account temporal variations of the detector configuration and data-taking conditions.

    Selection requirements are optimized using the figure of merit, FOM$= \frac{S}{\sqrt{S+B}}$, where the number of signal ($S$) and background ($B$) events are determined from simulation.
    The $\mathcal{B}[h_b(2P)\to\gamma\chi_{bJ}(1P)]$ values are assumed to be 1.4\%, 4.0\%, and 5.0\% for $J = 0, 1,$ and 2, respectively \cite{guo2016}. 
    The selection requirements are summarized in Table \ref{tab:selection}.
    Charged tracks must have transverse momentum above 50 MeV/c and originate from a cylindrical region of length 6.0 cm along the beam axis and radius 1.0 cm in the transverse plane, centered on the $e^{+}e^{-}$ interaction point.
    Moreover, we only retain tracks that are in the CDC geometric acceptance and are associated with more than 20 hits in this sub-detector.
    We select events with exactly four tracks.
    Muon candidates are identified by requiring 
    $\mathcal{P}_{\mu} = \frac{\mathcal{L}_{\mu}}{\mathcal{L}_{\mu} + \mathcal{L}_{\pi} + \mathcal{L}_K} >  0.8,$ where the likelihood 
    $\mathcal{L}_i, \ i=\mu, \pi, K$, is assigned based on the range of the charged particle extrapolated from the CDC through KLM and on deviation of hits from the extrapolated track \cite{Abashian:2002bd}.
    For pion candidates, we apply an electron veto $\mathcal{P}_e<0.4$, where $\mathcal{P}_e$ is a similar likelihood ratio based on CDC, ACC, and ECL information \cite{Hanagaki:2001fz}.
    To suppress the residual background from photon conversion, we apply a requirement on the opening angle between the pions in the laboratory frame, $\alpha_{\pi\pi}>11.7$°. 
    We apply a requirement on the $\mu^+\mu^-$ mass, $9.0 < M_{\mu\mu} < \qty{9.8}{GeV/c^2}$, which retains 96\% of the signal events. 
    
\begin{figure}[h!]
        \centering
        \includegraphics[width=0.48\textwidth]{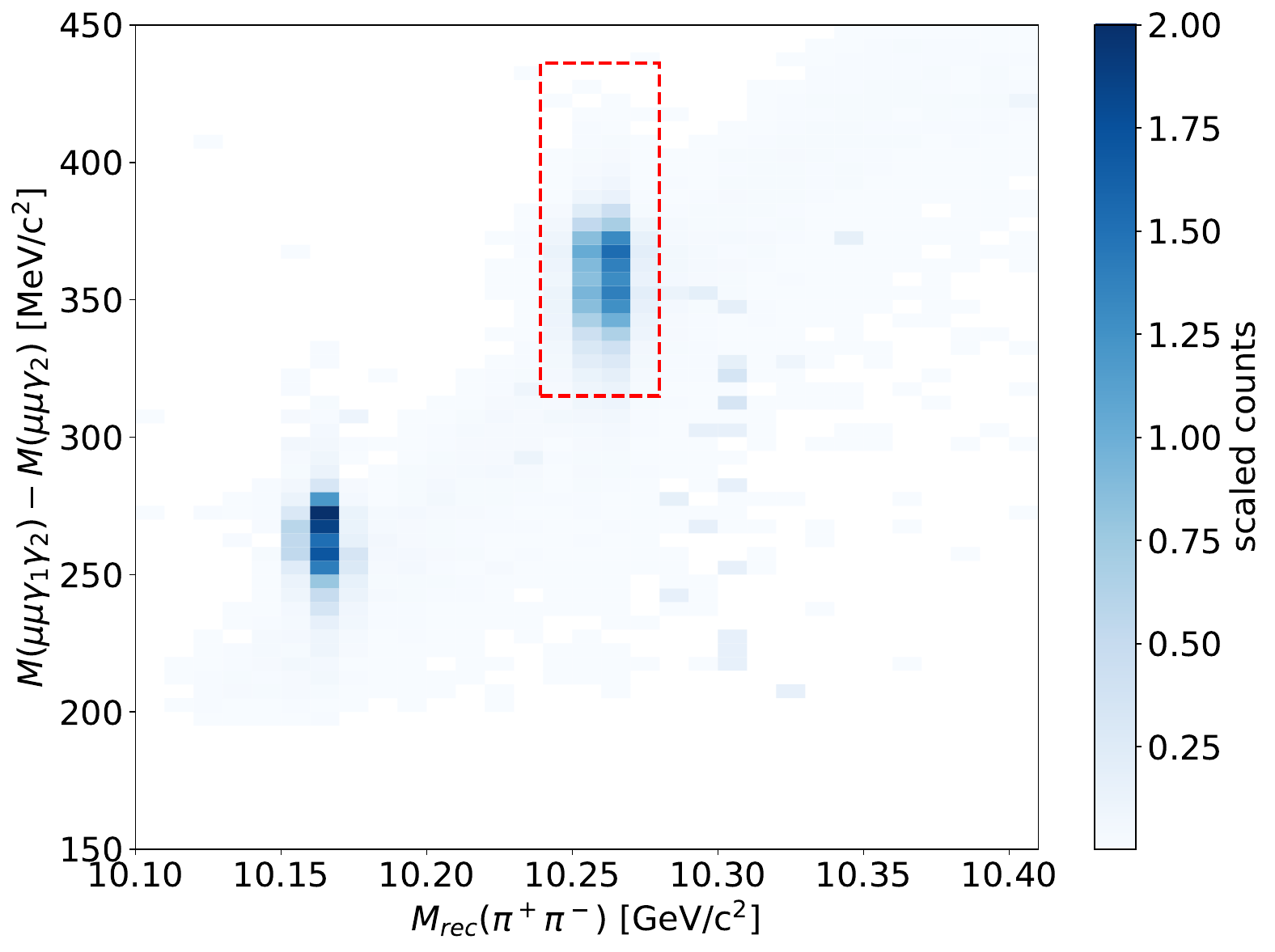}
        \caption{Two-dimensional distribution of the simulated events. The signal region is the rectangle defined by the conditions $10.239<M_{\textrm{rec}}(\pi^+\pi^-)<$ \qty{10.280}{GeV/\textit{c}^2} and $315<M(\mu\mu\gamma_1\gamma_2)-M(\mu\mu\gamma_2)<$ \qty{436}{MeV/\textit{c}^2}. The bump outside of this region corresponds to the simulated $\Upsilon(10860)\to\pi^+\pi^-\Upsilon_{\textrm{J}}(1D)\to\pi^+\pi^-\gamma\gamma\Upsilon(1S)$ transitions.}\label{fig:Mrecpipi}
    \end{figure}

    Photons are detected as ECL clusters without associated charged particles.
    We require the energies of the $\gamma_1$ and $\gamma_2$ candidates in the laboratory frame to exceed 267 MeV and 305 MeV, respectively.
    We veto the $\pi^0$ background, mostly given by the process $\Upsilon(10860)\to\Upsilon(2S)\pi^0\pi^0$, by discarding events with one or more candidates for which the di-photon invariant mass $M(\gamma_1\gamma_2)$ lies in a region of 20 MeV around the known $\pi^0$ mass.
    Similarly, we veto the $\eta \to \gamma\gamma$ decay requiring $|M(\gamma_1\gamma_2)-m(\eta)| > \qty{33.4}{MeV/c^2}$ to suppress the background coming from $h_b(2P)\to\eta\left[\gamma\gamma\right] \Upsilon(1S)\left[\mu^+\mu^-\right]$.
    Finally, we perform a 4C kinematic fit \cite{List:88030}, constraining the total four-momentum of the final state particles to the four-momentum of the initial state and requiring corresponding $p$-value to exceed $10^{-5}$. If multiple candidates are found, the one with the highest $p$-value is chosen.
    The signal region is defined as $10.239<M_{\textrm{rec}}(\pi^+\pi^-)<$ \qty{10.280}{GeV/\textit{c}^2} and $315<M(\mu\mu\gamma_1\gamma_2)-M(\mu\mu\gamma_2)<$ \qty{436}{MeV/\textit{c}^2}. The efficiencies of the selection requirements, summarized in Table 1, are $5.2\%, 10.9\%, 10.7\%$ for the $J = 0, 1, 2$ channels, respectively.
    Figure \ref{fig:Mrecpipi} shows the 2D distribution of simulated events and the signal region.

    \begin{table}[b]
    \caption{\label{tab:selection}
    Summary of the applied selection criteria.}
    \begin{ruledtabular}
    \begin{tabular}{ll}
    \textrm{Variable}&
    \textrm{Value}
    \\
    \colrule
    $M_{\mu\mu}$ & $[9.0, 9.8]$ \unit{GeV/\textit{c}^2} \\
    $p_T(\pi)$ & $ > 50$ \unit{MeV/\textit{c}}\\
    $E(\gamma_1)$ & $ > 267$ \unit{MeV}\\
    $E(\gamma_2)$ & $ > 305$ \unit{MeV}\\
    $\alpha_{\pi\pi}$  & $ > 11.7$°\\
    $|M(\gamma_1\gamma_2)-m(\pi^0)|$  & $> 10.0$ \unit{MeV/\textit{c}^2}\\
    $|M(\gamma_1\gamma_2)-m(\eta)|$  & $> 33.4$ \unit{MeV/\textit{c}^2}\\
    4C kinematic fit $p$-value & $ > 10^{-5}$\\
    $M_{\textrm{rec}}(\pi^+\pi^-)$ & $[10.239, 10.280]$ \unit{GeV/\textit{c}^2}\\
    $M(\mu\mu\gamma_1\gamma_2)-M(\mu\mu\gamma_2)$ & $[315, 436]$ \unit{MeV/\textit{c}^2}\\
    \end{tabular}
    \end{ruledtabular}
    \end{table}
    We observe a few background events in the simulated sample. A majority of them comes from the process $e^+e^-\to\pi^+\pi^-\pi^0\chi_{b1}(1P)$, where the three pions are produced either directly or by the decay of an intermediate $\omega$ meson. 
    This background, with an expected yield of 0.92 events, is irreducible, but the corresponding branching fraction has been measured \cite{Belle:2018izj} and can be used to estimate the expected number of events from simulation.
    Another source of background is $e^+e^- \to \mu^+\mu^-  3\gamma$, where one of the photons is converted in the detector. This sample yields 0.1 events.
    The last source of background is $h_b(2P)\to\eta\left[\gamma\gamma\right]\Upsilon(1S)\left[\mu^+\mu^-\right]$ with branching fraction $(7.1^{3.7}_{3.2}\pm0.8)\times10^{-3}$ \cite{Belle:2024qmw}, which is reconstructed with $0.05\%$ efficiency after applying the $\eta\to\gamma\gamma$ veto. From this source we retain 0.02 events.



\begin{figure}[h!]
        \includegraphics[width=0.48\textwidth]{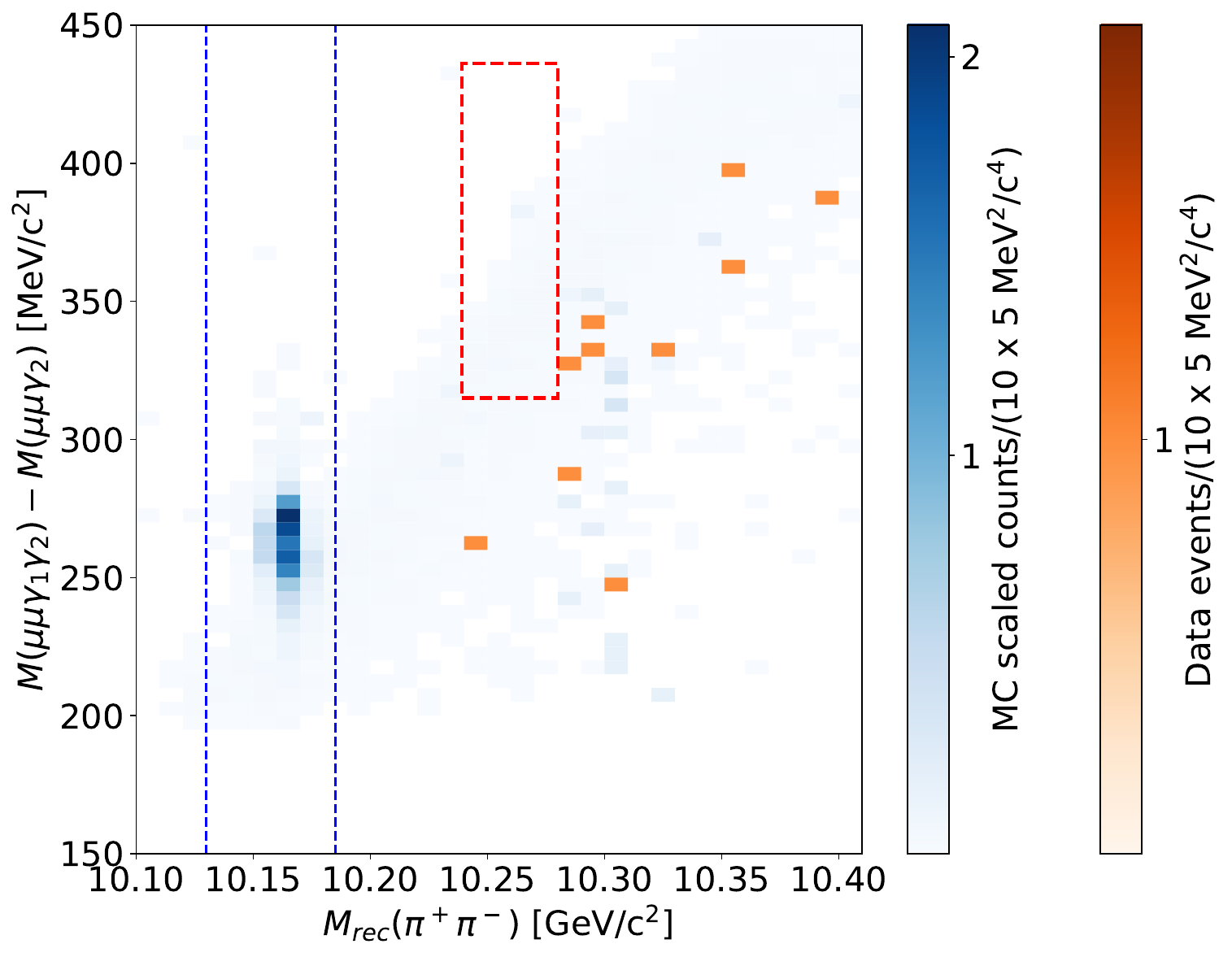}
        \caption{Two-dimensional distribution of experimental data and expected background. The red dashed rectangle defines the signal region. The blue dashed lines enclose the blind region $10.130<M_{\textrm{rec}}(\pi^+\pi^-)<10.185$ GeV/$c^2$ where the decay $\Upsilon(10860)\to\pi^+\pi^-\Upsilon_J(1D)\to\pi^+\pi^-\gamma\gamma\Upsilon(1S)$ may be observed.}
        \label{fig:fitvar}
    \end{figure}

    The two-dimensional distribution in $M_{\textrm{rec}}(\pi^+\pi^-)$ and $M(\mu\mu\gamma_1\gamma_2)-M(\mu\mu\gamma_2)$ for the candidates selected in data is shown in Fig.\ref{fig:fitvar}. 
    The region $10.130 < M_{\textrm{rec}}(\pi^+\pi^-) <$ \qty{10.185}{GeV/\textit{c}^2} is kept blind for it may contain events from the yet unobserved $e^+e^- \to \Upsilon_J(1D)\pi^+\pi^- \to \Upsilon(1S)\gamma\gamma\pi^+\pi^-$ cascade, which is currently under study at Belle.
    In the signal region, the observed count is zero on experimental data,
    while the background count is 1.04 in simulation.
    
    The expected number of background events is a key input for the estimation of the upper limits on the signal branching fractions. The correction factor and systematic uncertainty on the background counts are extracted from the data/MC comparison in the $M_{\textrm{rec}}(\pi^+\pi^-)$ sidebands.
    This comparison is done after applying a looser selection in order to increase statistics in the sidebands.
    With this selection we drop the 4C kinematic fit and lower the photon energy requirement to \qty{150}{MeV} for both $\gamma_1$ and $\gamma_2$.
    We define the three sidebands [10.000, 10.130], [10.185, 10.239], [10.280, 10.500] \unit{GeV/\textit{c}^2}.
    In the combined sidebands we measure $1466^{+39}_{-38}$ counts on data and $1444 \pm 45$ on simulation.
    The definition of the sidebands and the measured number of events, alongside with the data/MC ratios, are reported in Table \ref{tab:sidebands_counts}.
    \begin{table}[h!]
    \caption{\label{tab:sidebands_counts}
    Counted number of events in the sidebands of the $M_{\textrm{rec}}(\pi^+\pi^-)$ variable. The uncertainties include systematic contributions and are propagated by summing in quadrature.}
    \begin{ruledtabular}
    \begin{tabular}{lccc}
    \textrm{Sideband (GeV)}&
    \textrm{MC}&
    \textrm{Data}&
    \textrm{Data/MC}\\
    \colrule
$[10.000, 10.130]$ & $ 404^{+12}_{-11}$ & $ 390^{+21}_{-20}$ & $ 0.97 \pm 0.06$ \\
$[10.185, 10.239]$ & $ 98 \pm 12$ & $ 124^{+12}_{-11}$ & $ 1.26^{+0.20}_{-0.19}$ \\
$[10.280, 10.500]$  & $942\pm 25$      & $952^{+32}_{-31}$ & $ 1.01\pm0.04$ \\
All  & $ 1444 \pm 45$    & $ 1466^{+39}_{-38}$ & $ 1.02\pm0.04$ \\ 
    \end{tabular}
    \end{ruledtabular}
    \end{table}
    Based on the results from the combined sidebands, we find that the simulation correctly describes the background within a $\pm 4\%$ systematic uncertainty. 
    The expected number of background events in the signal region is $1.06^{+0.29}_{-0.20}$, where the uncertainty is both due to the correction factor and limited statistics in the simulated sample.

    In order to validate the analysis procedure and to estimate the systematic uncertainty in the efficiency, we apply a similar selection to a different data set taken at the $\Upsilon(3S)$ resonance. 
    The data set corresponds to \qty{2.9}{fb^{-1}} taken at the resonance peak plus \qty{0.3}{fb^{-1}} taken in an off-resonance scan. We reconstruct the cascade decays
    \begin{eqnarray*}\label{eq:cc_process}
        \Upsilon(3S) \rightarrow \gamma \chi_{bJ}(2P) \rightarrow \gamma\gamma\Upsilon(2S)\rightarrow\gamma\gamma\pi^+\pi^-\Upsilon(1S) \\ \rightarrow\gamma\gamma\pi^+\pi^-\mu^+\mu^-
    \end{eqnarray*}
    with $J$ = 1, 2. Concerning the topology, the main difference with respect to the signal channels is that the radiated photons are softer, however their energy is of the same order of magnitude ($\sim$ \qty{100}{MeV}).
    The dipion transition $\Upsilon(2S)\rightarrow\pi^+\pi^-\Upsilon(1S)$ ensures that pions have low momentum as for the signal cascade.
    We apply slightly different photon energy requirements than those described in Table \ref{tab:selection}.
    The photon energy in the lab frame is required to be higher than \qty{60}{MeV}.
    Photons with $80 < E_{\mathrm{c.m.}} <$ \qty{150}{MeV} are selected to reconstruct the $\Upsilon(3S)\to\gamma\chi_{bJ}(2P)$ decays, while photons with $160 < E_{\mathrm{c.m.}} <$ \qty{280}{MeV} are associated with the $\chi_{bJ}(2P)\to\gamma\Upsilon(2S)$ processes.
    A stringent requirement $9.770 < M_{\textrm{rec}}(\pi^+\pi^-) <$ \qty{9.810}{GeV/\textit{c}^2} is applied as for the signal channel. The $\pi^0\to\gamma\gamma$ is vetoed analogously to the signal selection, while the veto for $\eta\to\gamma\gamma$ is not applied.
    Due to the presence of multiple candidates, we perform a 4C kinematic fit and select the one with highest $p$-value.

\begin{figure}[h!]
        \centering
        \includegraphics[width=0.48\textwidth]{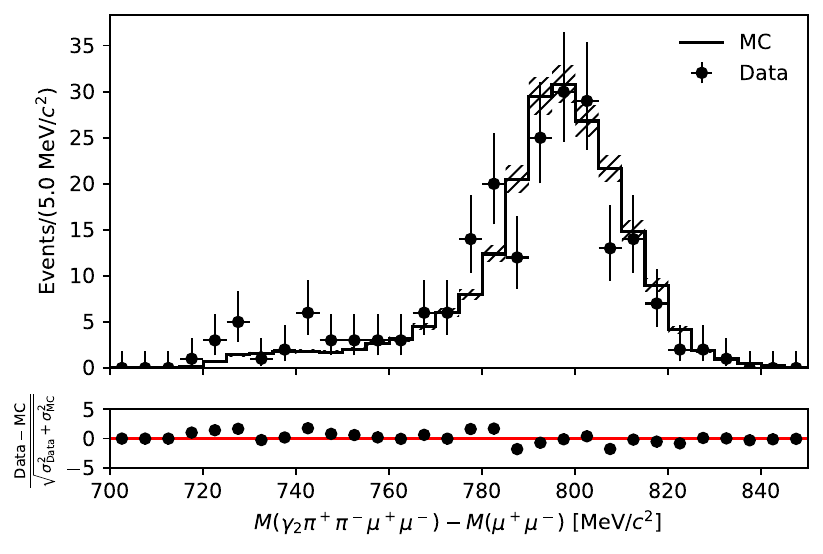}
        \caption{Distribution of the variable used to estimate the systematic uncertainty due to event selection. The error bars on data correspond to the Poisson 68\% confidence interval. The hatches on the MC histogram include the Poisson uncertainty and the systematic uncertainties propagated from the known branching fractions.}\label{fig:fit_cc}
    \end{figure}

    In Fig.\ref{fig:fit_cc} we show the data/MC comparison on the $M(\mu\mu\pi\pi\gamma_2)−M(\mu\mu)$ variable, where $\gamma_2$ is the photon radiated by the $\chi_{bJ}(2P)$ states. 
    The two distributions are background-free and agree well. 
    We count the number of events in the region $700 < M(\mu\mu\pi\pi\gamma_2)−M(\mu\mu) < 850$ MeV/$c^2$.
    We measure $N_{\textrm{Data}}=211^{+15}_{-14}$, $N_{\textrm{MC}}=209 \pm 24$, $N_{\textrm{Data}}/N_{\textrm{MC}} = 1.01 \pm 0.14$. The uncertainty on $N_{\textrm{MC}}$ includes systematic contributions from known branching fractions. Thus, we estimate an overall 14\% uncertainty on the efficiency estimation due to event selection.

    All the sources of systematic uncertainty and their values are summarized in Table \ref{tab:systematics}.
    \begin{table}[h!]
    \caption{\label{tab:systematics}%
    Summary of the systematic uncertainties contributing to the observed upper limits of the signal branching fractions.}
    \begin{ruledtabular}
    \begin{tabular}{ll}
    \textrm{Source}&
    \textrm{Uncertainty}\\
    \colrule
    Luminosity & $\pm$\qty{1.70}{fb^{-1}} \\
    $\sigma(e^+e^-\to h_b(2P)\pi^+\pi^-)$ & $\pm$\qty{360}{fb}  \\
    $\mathcal{B}[\chi_{b0}(1P)\to\gamma\Upsilon(1S)]$ & $\pm 0.27\times10^{-2}$ \\
    $\mathcal{B}[\chi_{b1}(1P)\to\gamma\Upsilon(1S)]$ & $\pm 2.0\times10^{-2}$ \\
    $\mathcal{B}[\chi_{b2}(1P)\to\gamma\Upsilon(1S)]$ & $\pm 1.0\times10^{-2}$ \\
    $\mathcal{B}[\Upsilon(1S)\to\mu^+\mu^-]$ & $\pm 0.05\times10^{-2}$ \\
    Efficiency & \quad \\
    \quad $h_b(2P)\to\gamma\chi_{b2}(1P)$ & $\pm 1.50\times10^{-2}$ \\
    \quad $h_b(2P)\to\gamma\chi_{b1}(1P)$ & $\pm 1.53\times10^{-2}$ \\
    \quad $h_b(2P)\to\gamma\chi_{b0}(1P)$ & $\pm 7.28\times10^{-3}$ \\
    Background counts & $\pm 4.24\times10^{-2}$\\
    \end{tabular}
    \end{ruledtabular}
    \end{table}

    We estimate the upper limits on the signal branching fractions by generating pseudo-experiments and applying the Feldman-Cousins method \cite{Feldman:1997qc} for the construction of the confidence belts. The number of pseudo-events in the $M(\mu\mu\gamma_1\gamma_2)-M(\mu\mu\gamma_2)$ variable for each channel $h_b(2P)\to\gamma\chi_{bJ}(1P)$, $J$ = 0,1,2, is extracted from a Poisson distribution with mean defined as
    \begin{equation*}
        \nu_{\textrm{sig}}(J) = L \, \sigma_{h_b\pi\pi} \, \mathcal{B}(J) \, \varepsilon_{J} \, \mathcal{B}_{\textrm{sig}}(J)
    \end{equation*}
    where $L$ is the luminosity at $\sqrt{s}=$ \qty{10.866}{GeV}, $\sigma_{h_b\pi\pi}=\sigma(e^+e^-\to h_b(2P)\pi^+\pi^-)$ \cite{Belle:2015tbu}, $\mathcal{B}(J)=\mathcal{B}[\chi_{bJ}(1P)\to\gamma\Upsilon(1S)]\times\mathcal{B}[\Upsilon(1S)\to\mu\mu]$ \cite{ParticleDataGroup:2022pth}, $\varepsilon_{J}$ is the selection efficiency and $\mathcal{B}_{\textrm{sig}}(J)=\mathcal{B}[h_b(2P)\to\gamma\chi_{bJ}(1P)]$.
    In order to include systematic uncertainties in the upper limit estimation, all the quantities except $\mathcal{B}_{\textrm{sig}}(J)$ are sampled from a Gaussian distribution with mean equal to the expected value and standard deviation equal to its uncertainty.
    The mean number of background events is $\nu_\textrm{b} = N_\textrm{b} s f$, where $N_\textrm{b}$ is sampled from a Poisson distribution with mean equal to the unweighted MC counts, $s$ is the mean weight, and $f$ is the correction factor, sampled from a Gaussian distribution with mean $1.02$ and standard deviation $0.04$ (see Table \ref{tab:sidebands_counts}).
    



    The upper limits at 90\% confidence level (CL) are estimated by using the Feldman-Cousins method, where the theoretical Poisson distribution is replaced by the distribution of counts generated in the pseudo-experiments.
    The results for the investigated channels are summarized in Table \ref{tab:results}.

    \begin{table}[h!]
    \caption{\label{tab:results}%
    Observed upper limits at 90\% CL for the branching fractions of the investigated transitions.}
    \begin{ruledtabular}
    \begin{tabular}{ll}
    \textrm{Channel}&
    \textrm{$\mathcal{B}$}\\
    \colrule
    $h_b(2P)\to\gamma\chi_{b2}(1P)$ & $ < 1.3 \times 10^{-2}$ \\
    $h_b(2P)\to\gamma\chi_{b1}(1P)$ & $ < 5.4 \times 10^{-3}$\\
    $h_b(2P)\to\gamma\chi_{b0}(1P)$ & $ < 2.7 \times 10^{-1}$    
    \end{tabular}
    \end{ruledtabular}
    \end{table}

In conclusion, we performed the first search for hindered M1 transitions between P-wave states of the  bottomonium system. Using the Belle data sample of \qty{121.4}{fb^{-1}} collected at $\sqrt{s}=$ \qty{10.866}{GeV}, we set 90\% CL upper limits for the branching fractions $\mathcal{B}[h_b(2P)\to\gamma\chi_{bJ}(1P)], \ J=0,1,2$:
\begin{flalign*}
\mathcal{B}[h_b(2P)\to\gamma\chi_{b0}(1P)] < 2.7 \times 10^{-1}, \\
\mathcal{B}[h_b(2P)\to\gamma\chi_{b1}(1P)] < 5.4 \times 10^{-3} \ \textrm{and} \\
\mathcal{B}[h_b(2P)\to\gamma\chi_{b2}(1P)] < 1.3 \times 10^{-2}.
\end{flalign*}
The upper limits are consistent with the Relativized Quark Model expectations \cite{Godfrey:1985xj} and help to constrain parameters of the coupled-channel model \cite{guo2016}. 

This work, based on data collected using the Belle detector, which was
operated until June 2010, was supported by 
the Ministry of Education, Culture, Sports, Science, and
Technology (MEXT) of Japan, the Japan Society for the 
Promotion of Science (JSPS), and the Tau-Lepton Physics 
Research Center of Nagoya University; 
the Australian Research Council including grants
DP210101900, 
DP210102831, 
DE220100462, 
LE210100098, 
LE230100085; 
Austrian Federal Ministry of Education, Science and Research (FWF) and
FWF Austrian Science Fund No.~P~31361-N36;
National Key R\&D Program of China under Contract No.~2022YFA1601903,
National Natural Science Foundation of China and research grants
No.~11575017,
No.~11761141009, 
No.~11705209, 
No.~11975076, 
No.~12135005, 
No.~12150004, 
No.~12161141008, 
and
No.~12175041, 
and Shandong Provincial Natural Science Foundation Project ZR2022JQ02;
the Czech Science Foundation Grant No. 22-18469S;
Horizon 2020 ERC Advanced Grant No.~884719 and ERC Starting Grant No.~947006 ``InterLeptons'' (European Union);
the Carl Zeiss Foundation, the Deutsche Forschungsgemeinschaft, the
Excellence Cluster Universe, and the VolkswagenStiftung;
the Department of Atomic Energy (Project Identification No. RTI 4002), the Department of Science and Technology of India,
and the UPES (India) SEED finding programs Nos. UPES/R\&D-SEED-INFRA/17052023/01 and UPES/R\&D-SOE/20062022/06; 
the Istituto Nazionale di Fisica Nucleare of Italy; 
National Research Foundation (NRF) of Korea Grant
Nos.~2016R1\-D1A1B\-02012900, 2018R1\-A2B\-3003643,
2018R1\-A6A1A\-06024970, RS\-2022\-00197659,
2019R1\-I1A3A\-01058933, 2021R1\-A6A1A\-03043957,
2021R1\-F1A\-1060423, 2021R1\-F1A\-1064008, 2022R1\-A2C\-1003993;
Radiation Science Research Institute, Foreign Large-size Research Facility Application Supporting project, the Global Science Experimental Data Hub Center of the Korea Institute of Science and Technology Information and KREONET/GLORIAD;
the Polish Ministry of Science and Higher Education and 
the National Science Center;
the Ministry of Science and Higher Education of the Russian Federation
and the HSE University Basic Research Program, Moscow; 
University of Tabuk research grants
S-1440-0321, S-0256-1438, and S-0280-1439 (Saudi Arabia);
the Slovenian Research Agency Grant Nos. J1-9124 and P1-0135;
Ikerbasque, Basque Foundation for Science, and the State Agency for Research
of the Spanish Ministry of Science and Innovation through Grant No. PID2022-136510NB-C33 (Spain);
the Swiss National Science Foundation; 
the Ministry of Education and the National Science and Technology Council of Taiwan;
and the United States Department of Energy and the National Science Foundation.
These acknowledgements are not to be interpreted as an endorsement of any
statement made by any of our institutes, funding agencies, governments, or
their representatives.
We thank the KEKB group for the excellent operation of the
accelerator; the KEK cryogenics group for the efficient
operation of the solenoid; and the KEK computer group and the Pacific Northwest National
Laboratory (PNNL) Environmental Molecular Sciences Laboratory (EMSL)
computing group for strong computing support; and the National
Institute of Informatics, and Science Information NETwork 6 (SINET6) for
valuable network support.

\nocite{*}

\bibliography{apssamp}

\end{document}